\documentclass[sigconf, balance]{acmart}

\usepackage[utf8]{inputenc}
\usepackage[T1]{fontenc}

\usepackage{multirow}

\usepackage{graphicx}
\usepackage{subcaption}		% composed images
\usepackage{tikz}				% Figure diagrams
\usetikzlibrary{shapes, arrows, chains, calc, positioning, automata, bending}
\usepackage{pgfplots}		% Low layer plotting and drawing 
\pgfplotsset{compat=newest} % Allows to place the legend below the plot
\usepgfplotslibrary{units}	% Allows to enter the units nicely

\pgfdeclarelayer{marx}
\pgfsetlayers{main,marx}
\providecommand{\cmark}[2][]{%
	\begin{pgfonlayer}{marx}
		\node [nmark] at (c#2#1) {#2};
	\end{pgfonlayer}{marx}
} 
\providecommand{\cmark}[2][]{\relax} 
\colorlet{lcfree}{green}
\colorlet{lcnorm}{blue}
\colorlet{lccong}{red}

\usepackage{multirow}	% Allow multirow grouping
\usepackage{booktabs}

\usepackage{enumitem}
\usepackage{microtype}

\copyrightyear{2025}
\acmYear{2025}
\setcopyright{cc}
\setcctype{by}
\acmConference[IOT 2025]{The 15th International Conference on the Internet of Things}{November 18--21, 2025}{Vienna, Austria}
\acmBooktitle{The 15th International Conference on the Internet of Things (IOT 2025), November 18--21, 2025, Vienna, Austria}
\acmPrice{}
\acmDOI{10.1145/3770501.3770524}
\acmISBN{979-8-4007-1595-2/25/11}

\begin{document}
	
	\title{On the cybersecurity of {LoRaWAN}-based system: a Smart-Lighting case study}
	
	\author{Florian Hofer}
	\orcid{0000-0001-8336-0834}
	\affiliation{%
		\institution{Free University of Bolzano-Bozen}
		\city{Bolzano}
		\state{}
		\country{Italy}
	}
	\email{florian.hofer@unibz.it}
	
	\author{Barbara Russo}
	\orcid{0000-0003-3737-9264}
	\affiliation{%
		\institution{Free University of Bolzano-Bozen}
		\city{Bolzano}
		\country{Italy}
	}
	\email{barbara.russo@unibz.it}
	
	\renewcommand{\shortauthors}{F. Hofer and B. Russo}
	
	\begin{abstract}
		Cyber-physical systems and the Internet of Things (IoT) are key technologies in the Industry 4.0 vision. They incorporate sensors and actuators to interact with the physical environment. However, when creating and interconnecting components to form a heterogeneous smart systems architecture, these face challenges in cybersecurity. This paper presents an experimental investigation of architectural configurations for a LoRaWAN-based Smart-Lighting project, aimed at verifying and improving the system's robustness against attacks. We assess the system's robustness in a series of iterative experiments conducted both in-vitro and on-site. The results show that most attacks on a LoRaWAN network are unsuccessful, also highlighting unresolved issues with the installed products. The most successful attacks are high-power jamming attacks within a few meters of the target, which, in the case of gateways, can be mitigated through gateway redundancy.  
	\end{abstract}

	
	
	
	\maketitle
	
	%----------------------------------------
	% Set defaults for this section's plots
	%----------------------------------------
	\pgfplotsset{every axis/.append style={%
			y unit=ms,
			width=0.99\linewidth, % Scale the plot to \linewidth
			grid=major, 
			grid style={dashed, gray!30},
			xlabel=Experiment run, % Set the labels
			ylabel=RTT,
			x tick label style={rotate=90,anchor=east},
			only marks, 
			y filter/.expression={y==0 ? +inf : y},
			table/header=false,
			domain=1:30%
		},
		cycle list={%
			{blue,mark=asterisk},
			{red, mark=+},
			{orange, mark=diamond},
			{green, mark=square}%
		}
	}
	
	\section{Introduction}
	
	Smart systems operate a network of intelligent devices of various makes and functions to make autonomous, decentralized decisions, requiring standardized interfacing and communication. This heterogeneity may result in inconsistencies, leaving a system susceptible to attacks that can exploit these flaws to eavesdrop on or degrade the value of an asset, resulting in both virtual and physical losses~\cite {AshibaniMahmoud2017, Hofer2018}. Consequently, security must be built into modern architectural designs (\textit{security by design}). 
	
	However, in a multi-domain context that spans the cloud, edge, and physical world, where definitions and analytical models vary between application-relevant functional models, this is a complex process to accomplish. As a result, security has often been overlooked. Moreover, existing architectural models are typically troubled by simplifications and assumptions from the "offline" secure world. Thus, it is necessary to assess multi-domain vulnerabilities in systems, propose strategic and preventive countermeasures~\cite{Lezzietal2018}, or determine corrective mitigation means that could reduce or eliminate a vulnerability~\cite{Moteff2005}.
	
	This paper presents an iterative, experimental analysis of such a multi-domain system. We explore the network configuration and communication parameters of a decentralized \textit{Smart-Lighting} architecture (SLA) and assess its cybersecurity in-vitro and in an on-site case study. Smart-Lighting deployments are particularly susceptible to attacks due to multiple constraints such as bandwidth limitations and coexistence with other networks on an open band~\cite{Augustinetal2016}. Challenges like its widespread nature and the vast scale of such a network expose it to a higher risk of tampering and attacks. Based on the results of a previous offline analysis of the targeted SLA~\cite{HoferRusso2020}, we thus continue with an experimental assessment of the identified weaknesses and discuss potential future actions. Our contributions include a method for the experimental evaluation of the resilience of smart infrastructure to attacks, as well as an evaluation of a deployed architecture in an urban setting with and without the presence of an attacker.
	
	We organized the rest of the paper as follows. Section~\ref{sec:relwork} and~\ref{sec:method} present related work and our experimentation method, setup, and configuration. In Section~\ref{sec:exp} and ~\ref{sec:discussion}, we report and discuss execution results and consequences of the experiments. Section~\ref{sec:conclusions} concludes the paper.
	
	\section{Related work}
	\label{sec:relwork}
	
	Being a relatively recent technology, LoRaWAN has generated numerous articles discussing its limits and capabilities. Adelantado et al.~\cite{Adelantadoetal2017} focus on the limitations and application scenarios of such a low-throughput, high-range communication method. Their theoretical exploration concludes similarly to Augustin et al.'s~\cite{Augustinetal2016} experimental evaluation, which suggests that LoRaWAN suffers from coexistence and bandwidth limitations, making it an easy target for attackers. Recently, Šabić et al.~\cite{Sabicetal2024} have extensively summarized and reiterated these affirmations, adding that, since LoRaWAN devices are inexpensive, they are more likely to be targeted by such an attack. However, both of the former explore ALOHA-like throughput caps and the exponential collision behavior of the network, but fail to comment on the advantages of gateway redundancy, the modulation robustness, or general multi-gateway applications. Cited solutions and scenarios primarily focus on one-gateway analysis and approaches, overlooking the benefits of an inexpensive LoRaWan redundancy setup. Unlike them, Georgiou and Raza~\cite{GeorgiouRaza2017} discovered the same limitations in their stochastic analysis of a one-gateway infrastructure. However, they acknowledged that multiple gateways would change the behavior and throughput of the system. 
	
	Indeed, an in-depth analysis of Semtech's proprietary LoRa modulation revealed that existing models are inadequate, as noted by Afisiadis et al.~\cite{Afisiadisetal2020}. They assessed LoRa's performance in the presence of additive white Gaussian noise and interference from another LoRa user, or attacker, without assuming perfect signal alignment. Their numerical simulations validated the model, showing that existing interference models had overestimated transmission error rates. Likewise, an experimental evaluation of the protocol's collision and reconstruction management by Haxhibeqiri et al.~\cite{Haxhibeqirietal2017} highlighted its robustness. It showed how timing and signal strength are the primary factors that determine whether and which of the two colliding signals is received. Thus, they demonstrated LoRaWAN's better scalability and resilience to attacks than previously assumed. 
	
	Other research exists that has similarly explored the vulnerabilities of LoRaWAN networks. Yang et al.~\cite{Yangetal2018}, e.g., analyze and compare present vulnerabilities of LoRaWan 1.0.2 and perform proof of concept tests in a one-gateway in-vitro setup. The work highlights the progressive improvement of the standard and the weaknesses of a manually configured device. Torres et al.~\cite{Torresetal2022} performed tests on a single-gateway in-vitro setup in an attempt to detect typical physical layer vulnerabilities. Similarly, Aras et al.~\cite{Arasetal2017} and Martinez et al.~\cite{Martinezetal2019, Martinezetal2020} conducted multiple studies to assess the performance of a single-gateway LoRaWAN network under jamming conditions. However, these consider single gateway networks or even attackers that transmit according to specifications only.
	
	We will thus experimentally assess the network behavior, resilience, and jamming efficiency with multiple gateways in place.
	
	\section{Experiment setup and configuration}
	\label{sec:method}
	
	\begin{figure}[!t]
		\centering
		\begin{tikzpicture}[%
			>=triangle 60,              % Nice arrows; your taste may be different
			start chain=going below,    % General flow is top-to-bottom
			node distance=6mm and 60mm, % Global setup of box spacing
			every join/.style={norm},   % Default line type for connecting boxes
			% Parameters for scaling TikZ
			%thick,
			scale=0.6,
			every node/.style={transform shape}
			]
			% ------------------------------------------------- 
			% A few box styles 
			% <on chain> *and* <on grid> reduce the need for manual relative
			% positioning of nodes
			\tikzset{
				base/.style={draw, on chain, on grid, align=center, minimum height=4ex},
				proc/.style={base, rectangle, text width=8em},
				test/.style={base, diamond, aspect=2, text width=5em},
				term/.style={proc, rounded corners},
				coord/.style={coordinate, on chain, on grid, node distance=6mm and 25mm},
				% nmark node style is used for coordinate debugging marks
				nmark/.style={draw, cyan, circle, font={\sffamily\bfseries}},
				% -------------------------------------------------
				% Connector line styles for different parts of the diagram
				norm/.style={->, draw, lcnorm},
				free/.style={->, draw, lcfree},
				cong/.style={->, draw, lccong},
				it/.style={font={\small\itshape}}
			}
			% -------------------------------------------------
			% Start by placing the nodes
			\node [proc, densely dotted, it] (p0) {Test specifications};
			% Use join to connect a node to the previous one 
			\node [term, join] (p1) {Start experiments};
			\node [proc] 	   (p2) {Update test scripts};
			\node [proc, join] (l1)	{Run experiments group};
			\node [proc, join] (p3)	{Evaluate results};
			\node [test, join] (t1) {Test optimized?};
			\node [proc] 	   (p4) {Run selected configuration(s)};
			
			\node [proc, densely dotted, it, right=of p0, xshift=-20mm, yshift=1mm] (e1) {Experimental hardware evaluation};
			
			% Next column
			\node [proc, right=of p1] (p5) {Tuning of tests};
			\node [proc, join] (p8) {Run test group};
			\node [proc, join] (p6) {Result evaluation};
			\node [test, join] (t2) {Comparable results?};
			\node [test] 	   (t3) {Groups finished?};
			\node [proc] 	   (p7) {Online tests};
			\node [term, join] (p10) {End experiments};
			% -------------------------------------------------
			\node [coord, left=of p4] (c0)  {}; 
			\node [coord, right=of p2] (c1)  {};
			\node [coord, below=of c1] (c11)  {}; \cmark[1]{1}
			\node [coord, right=of p5] (c2)  {}; 
			\node [coord, above right=2mm of c2] (c21)  {}; \cmark[1]{2}
			% -------------------------------------------------
			% A couple of boxes have annotations
			\node [above=0mm of p2, it] {(In-Vitro)};
			\node [above=0mm of p5, it] {(On-Site)};
			% -------------------------------------------------
			% All the other connections come out of tests and need annotating
			% First, the straight north-south connections. In each case, we first
			% draw a path with a (consistently positioned) annotation node, then
			% We draw the arrow itself.
			\path (t1.south) to node [near start, xshift=1em] {$y$} (p4); 
			\draw [->,lcnorm] (t1.south) -- (p4);
			\path (t2.south) to node [near start, xshift=1em] {$y$} (t3); 
			\draw [->,lcnorm] (t2.south) -- (t3);
			\path (t3.south) to node [near start, xshift=1em] {$y$} (p7); 
			\draw [->,lcnorm] (t3.south) -- (p7);
			
			\draw [->,lcnorm] (p4.east) -- ++(15mm,0mm) |- (p5);
			\path (t1.east) -| node [very near start, yshift=-1em, xshift=2mm] {\shortstack{$n$\\\tiny(adjust parameters)}} (c1); 
			\draw [->,lcfree] (t1.east) -| (c1)  |- (p2);
			\path (t2.east) -| node [very near start, yshift=-1em, xshift=2mm] {\shortstack{$n$\\\tiny(adjust parameters)}} (c2); 
			\draw [->,lcfree] (t2.east) -| (c2) -- (p5); 
			\path (t3.west) -| node [very near start, yshift=-1em, xshift=2mm] {\shortstack{$n$\\\tiny(next group)}} (c1); 
			\draw [-,lcfree] (t3.west) -| (c1); 
			% -------------------------------------------------
			\draw [-,lcnorm, gray, densely dotted] (p1.south) -- (p2.north);
			\draw [->,lcnorm] (e1.west) -| ($(e1.west)+(-3mm,0mm)$) |-  (p1.east);
			
			% -------------------------------------------------
			% Loop arrows
			\draw [->, green!50!black] ($(l1.south east) +(2mm,1mm)$) arc (20:360:3mm);
			\draw [->, green!50!black] ($(p3.south east) +(2mm,1mm)$) arc (20:360:3mm);
			\draw [->, green!50!black] ($(p4.south east) +(2mm,1mm)$) arc (20:360:3mm);
			\draw [->, green!50!black] ($(p8.south east) +(2mm,1mm)$) arc (20:360:3mm);
			\draw [->, green!50!black] ($(p6.south east) +(2mm,1mm)$) arc (20:360:3mm);
			\draw [->, green!50!black] ($(p7.south east) +(2mm,1mm)$) arc (20:360:3mm);

			% -------------------------------------------------
			% Highlighting boxes 
			\draw[orange,thick,dotted] ($(p2.north west)+(-5mm,5mm)$) rectangle ($(p4.south east)+(5mm,-3mm)$);
			\draw[brown,thick,dotted]  ($(p5.north west)+(-5mm,5mm)$) rectangle ($(p7.north east)+(5mm,3mm)$);
			
			% -------------------------------------------------
			% 
			\draw [->, green!50!black] ($(p4.south west) +(0mm,-12mm)$) arc (20:360:3mm);
			\node [black,below left=10mm and 0mm of p4, it, anchor=north west] {\shortstack{denotes iterative tasks done once\\for each configuration combination}};
		\end{tikzpicture}
		\caption{Experiment (re)design and execution method}
		\label{fig:flowchart}
		\Description[Flowchart of the experimentation process]{The image displays a flowchart in two phases that iterates over multiple steps, first in-vitro then on-site, to help determine system configurations for the tests at hand} 
	\end{figure}
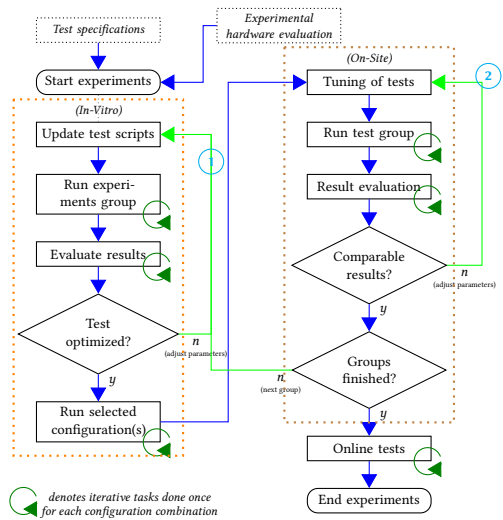
	
	We implement and execute experiments to measure the system's robustness through two test batch cycles between an in-vitro setup and the on-site deployed SLA. Through the latter pilot project, we will determine communication and architecture parameters that achieve a better, more secure operation before an extensive roll-out of the system.
	
	The gateway and test hardware and positions are the result of an experimental analysis aimed at exploiting the concave shape of the terrain to improve coverage~\cite{HoferKuen2022}, with a target count of 1500 devices and at least three gateways in reach per node, Fig.~\ref{fig:gw_locations}. Test configurations are further selected based on the SLA's hardware, software, as well as the results of a previously performed offline security analysis of the deployment~\cite{HoferRusso2020}. During each test batch, parameters are verified and tuned in in-vitro (Figure~\ref{fig:flowchart} left) and replicated and verified on-site (Figure~\ref{fig:flowchart} right). Secondly, each cycle, whether in-vitro or on-site, informs the next repetition by adjusting and redesigning the test method. 
	
	After the resilience and tuning tests, we conduct a set of online overnight tests that simulate the typical operational scenario. We will thus observe the change in communication performance with and without the induced interference of a possible attacker by comparing the expected with the registered traffic. Through subsequent iterative adjustments of network parameters, we aim to enhance this performance and terminate once we achieve a predetermined message coverage of at least 97\%.
	
	The offline security analysis of the deployed SLA identified the following possible vulnerabilities: resource constraint, large scale, wide distribution, software bugs, inconsistent protocols, incorrect device configurations, and inconsistent and incomplete specifications. Based on this outcome, we developed tests to evaluate the presence of such vulnerabilities in our SLA. We exploited four vulnerabilities $v_i$, as follows:
	
	\begin{enumerate}[leftmargin=2em,label={$v_{\alph*}$}]
		\item \textit{resource constraint}: refers to the limited resources in a network, such as radio channels, gateways, end devices, and servers, which can become bottlenecks. Targeted network attacks can deplete communication or energy resources, hindering network functionality. The limited availability of ether restricts LoRaWAN's transmission capacity and facilitates channel interference by attackers. Targeted tests will aim to exhaust these resources.
		
		\item  \textit{large scale}: refers to the amount of attacking surface. A large-scale network faces channel contention and coexistence issues, prompting the use of airtime management and transmission power control to reduce signal range and interference. These measures can inadvertently help attackers by providing multiple targets within reach, increasing their likelihood of success. Tests at different distances will assess the impact of such attacks.
		
		\item  \textit{wide distribution}: highlights the large scale of the system of systems, making seamless safety and security monitoring difficult. This distribution increases vulnerability to attacks, as unattended areas facilitate device tampering, interference, and destruction. Such conditions allow for unauthorized physical interactions that can alter measurements. Upcoming distance tests will assess whether a compromised node can cause disruption. They will include scenarios with fake 'join' communications and jamming attacks, both with and without stolen credentials, to explore the effects of device capture.
		
		\item  \textit{incorrect device configurations}: can lead to bottlenecks in LoRa's ether resources due to neglected channels. Similarly, improper gateway setups can facilitate preamble-based resource availability attacks. Human errors, such as inconsistent permission schemes, may enable attackers to access systems or cause failures. Resource exhaustion attacks can address these issues.
	\end{enumerate}
	
	During the experiments, we also observed whether the following vulnerabilities are present:
	\begin{enumerate}[leftmargin=2em,label={$v_{\alph*}$},start=5]
		\item  \textit{software bugs} can grant unauthorized access to systems and cause network malfunctions or erratic radio behavior. We will identify this vulnerability through tests on a node that has supposedly been maliciously captured. By conducting long-term tests with and without interference, we aim to analyze the system's behavior under stress and identify any erratic responses associated with this issue.
		
		\item  \textit{inconsistent protocols} also hold issues of unauthorized access to infrastructure and information, or may similarly cause network malfunctions—tests as above.
		
		\item  \textit{inconsistent and incomplete specifications} may arise due to multiple versions and adaptations of the Alliance's protocol~\cite{Lora2016, Lora2018, Lora2020-2}. Consequently, implementations might not fully adhere to expected behavior, leading to inconsistent transmissions or erratic radio behavior. We will verify this potential weakness through long-term tests.
	\end{enumerate}
	
	\subsection{Hardware and Software configuration}
	
	\begin{figure}[tb]
		\centering
		\includegraphics[width=0.7\linewidth]{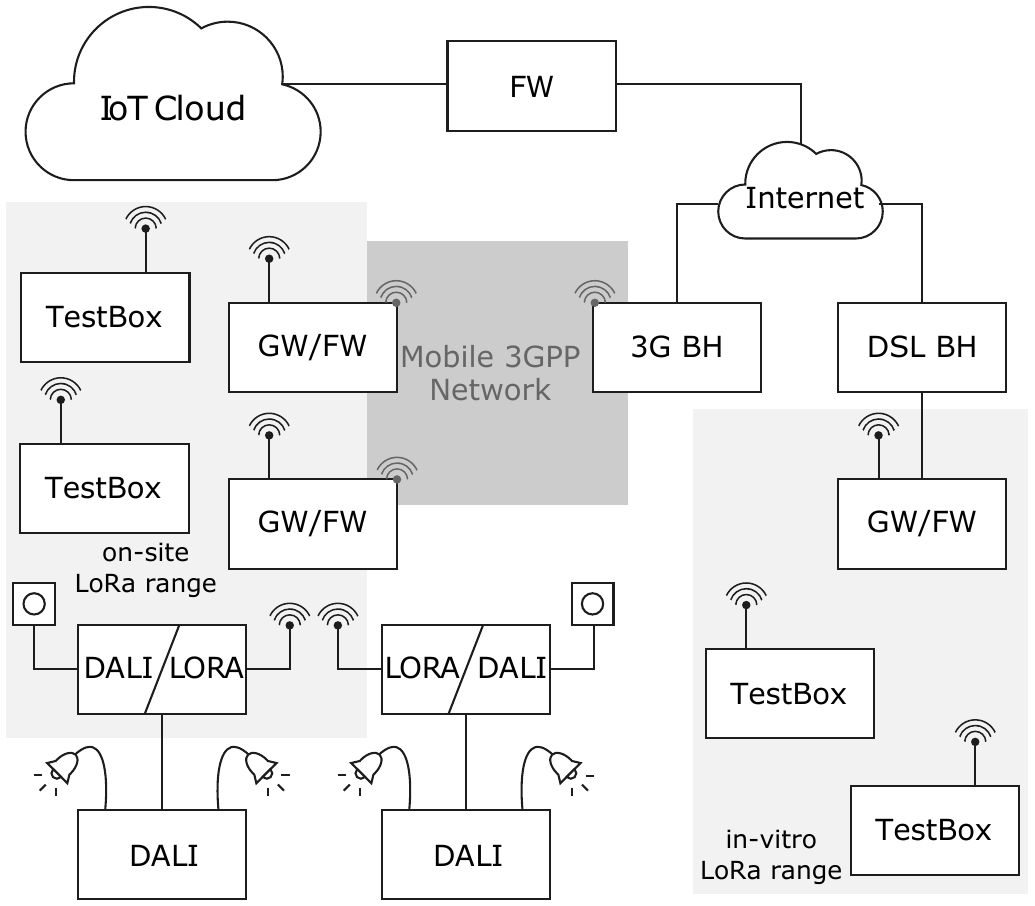}
		\caption{Experiment setup with backhaul (BH), gateway (GW), firewall (FW), and TestBox configurations.}
		\label{fig:communication}
		\Description[Graph of experiment setup]{This graph details components involved for the in-vitro and on-site network tests} 
	\end{figure}
	
	The pilot project mounts 81 lamp posts featuring different generations of Nordic Automation Systems (NAS) LoRaWAN to DALI Lighting controllers and six Wirnet Station 868 Kerlink gateways, as shown in Figure~\ref{fig:communication}. Each controller periodically sends confirmed status telegrams, unconfirmed usage statistics, and a confirmed boot information update at power-up. Confirmed messages require server acknowledgment and generate additional downlink traffic for every incoming message until received. Payloads for these uplinks range from 11 to 16 bytes, depending on the device profile, with message intervals varying from every 15 minutes to once daily. For the Lighting controller emulation, we choose these three data lengths: 1 byte (lower limit) for interference tests, 16 bytes in 15-minute intervals for lamp messages, and 51 bytes -- the maximum permitted data size for LoRaWAN packets using a repeater on the lowest speed\footnote{~we need to limit the data length as some tests use automatic adjustment}.
	
	Two test boxes synchronized through an ad-hoc Wi-Fi connection will autonomously execute tests both in-vitro and on-site. Each box contains five LoRaWAN-capable microcontrollers and a Raspberry Pi 3 B+ acting as an execution agent. These controllers are of two types~\cite{HoferKuen2022}: one mimics an on-site sensor performing a specific task tailored to the job it performs, i.e., an end node, and the other acts as an external agent to reproduce or interfere with the environment, i.e., a testing node or an attacker. The former, \emph{The Things Network Uno}, features onboard LoRaWAN connectivity and a ready-to-use library. Furthermore, its built-in PCB antenna reproduces the boxed design of the installed lighting controllers. The latter, \emph{MKRWan 1300}, features a 32-bit processor and direct radio access for unconstrained transmission\footnote{~Project repository and MCU code available at \url{https://flhofer.github.io/I4S/}}. 
	
	A LorIoT Osprey $7.x$ network server records the traffic data for analysis. As a performance metric, we use either the message round-trip time (RTT) for confirmed transmission or signal quality and success rates. The former combines transmission airtime to the Network server, protocol-required pauses and repeat intervals, and the response's airtime, which depends on the data rate and retransmission count, indirectly dependent on collisions and reception quality. Timestamped log files will capture outputs from the test boxes and server logs. 
	
	\subsection{LoRaWAN parameters}
	
	\begin{table}[t]
		\centering
		\renewcommand{\arraystretch}{1.1}
		\caption{Semtech's default EU868 LoRaWan parameters}
		\begin{tabular}{l p{0.28\linewidth} c p{5em}}
			\toprule
			& Frequency [MHz] & Duty Cycle & DR (SF)\\
			\midrule
			Tx Chn 1-3 & $868.1$, $868.3$, $868.5$ & 1\% & 0(12) to 5(7)\\
			Tx Chn 4-8 & $867.1$, $867.3$, $867.5$, $867.7$, $867.9$ & 0.1\%  & 0(12) to 5(7)\\
			Tx Chn 9 	& $868.3$ & 1\%  & 6(7)\\
			RX1 (GW Tx) & \multicolumn{3}{c}{same as Tx Chn 1-8}\\
			RX2 (GW Tx) & $869.525$ & 10\%  & 0(12)\\
			\hline
			Bandwidth & \multicolumn{3}{c}{125kHz, 250kHz for Tx Chn 9, @ CR 4/5}\\
			Tx Power[dBm]& \multicolumn{3}{c}{16 down-to 2 (index 0 to 7), default 14 (1)} \\
			\bottomrule
		\end{tabular}
		\label{tab:channels}
	\end{table}
	% $1\%$ for G, G1, and G4, $0.1\%$ for G2, and $10\%$ for the G3 sub-band
	
	To test the performance of a LoRaWAN Network, we must first gain a better understanding of its transmission behavior. LoRaWAN is a loosely coupled star-of-stars network based on the LoRa physical layer, which allows simultaneous wireless transmissions on one of the configured \textit{physical} channels, i.e., a preset frequency, to every gateway within reach. After each transmission, the end node switches to a new channel in a pseudo-random fashion. It transmits on the new channel (TX window), listens on the same channel (receive-window RX1), and finally listens to a second shared downlink channel (receive-window RX2). If the device is not battery-powered, it will listen to the latter whenever possible.
	
	A LoRa \textit{logical} transmission channel is a collection of modulation type, frequency, bandwidth, and data rate. For more robust communication, the more popular Chirp Spread Spectrum (CSS) modulation employs a code rate (CR) for forward error correction. It translates raw bits into symbols according to a spread factor (SF). The bigger the SF, the smaller the possible maximum data rate, but the more robust the transmission. At the same CR, the possible data rate (DR) is thus inversely proportional to the spread factor. CSS's particularity is its high channel selectivity, allowing simultaneous transmissions with different SFs. A recently introduced concept, Long Range Frequency Hopping Spread Spectrum (LR-FHSS), enables more robust uplink communication at the expense of lower data rates. However, there are no non-developer products yet on the market. 
	
	European LoRaWAN specifications define two ISM bands: a bandwidth- and power-restricted EU433 and the EU868, which device manufacturers most commonly use~\cite{Lora2020}. It further regulates band sharing via a transmission duty cycle. The protocol foresees a fixed CR of $4/5$ and a mechanism of adaptive data rate (ADR), which adjusts the set data rate, typically the network server, through downlink information when the signal quality is insufficient (less) or excellent (more). We can \emph{join} a network either via a pair of pre-distributed, manually uploaded session keys for the communication (ABP) or through an over-the-air activation (OTAA) procedure~cite{Lora2017-2}. The latter sends a message containing a small payload encrypted with a device-specific application key to request a new pair of session keys and communication configuration from a network (managing) server. 
	
	Each device must store multiple channel configurations. The specification for EU868 requires three default LoRaWAN physical channels (Tx Chn 1-3) that each device must support, in addition to up to 13 configurable channels in the $863-870MHz$ range. Semtech proposes the parameters in Table~\ref{tab:channels} totaling 49 \textit{logical channels}, i.e., the combination of the physical channel (frequency) and spread factor. For RX1, the device may use a DR offset w.r.t. uplink to reduce the risk of collisions between uplinks and downlinks. Other parameters exist that define the communication protocol. For example, the reception window duration and delay define when and how long the end nodes listen. Those are essential for battery-powered systems, as they affect power consumption. Similarly, the maximum retransmission counter, the retry delay, and the acknowledgment timeout govern the retransmission of messages that require confirmation. However, since these settings typically remain at their default values and do not significantly impact the radio exchange, we will keep them unchanged for simplicity.
	
	\section{Experimentation}
	\label{sec:exp}
	
	\subsection{Attacks for vulnerability exploitation}
	
	In this first batch, we address the vulnerabilities with directed tests. The end node sends 30 confirmed one, 16, or 51-byte messages starting with DR5 at LoRaWAN's power index $1$, and we observe a message's RTT. The simulation devices, meanwhile, attempt to disrupt the network using various strategies, including obstruction, interference, and resource exhaustion. Initially, we test the network's resilience to frequency jamming, where an attacker attempts to transmit on a resource—the ether—but does not know the exact transmission frequency, i.e., indirect interference. Next, we repeat the test assuming the attacker knows the LoRaWan frequencies, i.e., intentional obstruction. Finally, we test for different ways of causing resource exhaustion: flooding the channel with messages, flooding the gateway with messages, and flooding the network server with seemingly valid information.
	
	\begin{figure}[tb]
		\centering
		\begin{subfigure}[t]{0.30\textwidth}
			\centering
			\begin{tikzpicture}
				\begin{axis}[%
					legend style={at={(0.57,0.5)},
						anchor=south}
					]
					\addplot
					table[x index=0,y expr=\number\thisrowno{1} * \thisrowno{2},col sep=semicolon] 	{plotData/prelim_B3_SF12C8.csv}; 
					\addplot
					table[x index=0,y expr=\number\thisrowno{1} * \thisrowno{2},col sep=semicolon] 	{plotData/prelim_B3_SF12C5.csv}; 
					\addplot
					table[x index=0,y expr=\number\thisrowno{1} * \thisrowno{2},col sep=semicolon] 	{plotData/prelim_B3_SF7C8.csv}; 
					\addplot
					table[x index=0,y expr=\number\thisrowno{1} * \thisrowno{2},col sep=semicolon] 	{plotData/prelim_B3_SF7C5.csv};

					\legend{SF12 CR4/8, SF12 CR4/5,SF7 CR4/8,SF7 CR4/5}	
				\end{axis}
			\end{tikzpicture}
			\caption{Round-Trip time, end-node Tx 1 byte}
			\label{fig:two_B3tunea}			
		\end{subfigure}
		\begin{subfigure}[t]{0.30\textwidth}
			\centering
			\begin{tikzpicture}
				\begin{axis}[%
					ylabel=RSSI,
					y unit=dBm
					]
					\addplot
					table[x index=0,y expr=\number\thisrowno{10} * \thisrowno{2},col sep=semicolon] 	{plotData/prelim_B3_SF12C8.csv}; 
					\addplot
					table[x index=0,y expr=\number\thisrowno{10} * \thisrowno{2},col sep=semicolon] 	{plotData/prelim_B3_SF12C5.csv}; 
					\addplot
					table[x index=0,y expr=\number\thisrowno{10} * \thisrowno{2},col sep=semicolon] 	{plotData/prelim_B3_SF7C8.csv}; 
					\addplot
					table[x index=0,y expr=\number\thisrowno{10} * \thisrowno{2},col sep=semicolon] 	{plotData/prelim_B3_SF7C5.csv}; 
					
				\end{axis}
			\end{tikzpicture}
			\caption{Down-link receiver strength (RX1)}
			\label{fig:two_B3tuneb}
		\end{subfigure}
		\caption{LoRa indirect interference tests, in-vitro.}
		\label{fig:two_B3tune}
	\end{figure}
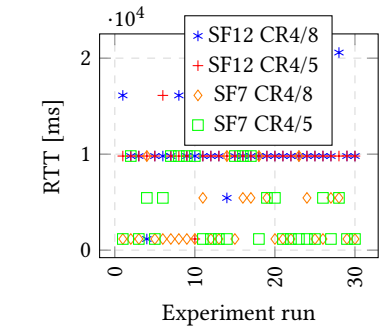
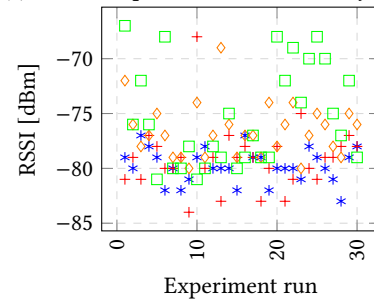
	
	\subsubsection{Tuning indirect interference}
	During the indirect interference tests, the end node operates only on the three default channels, $868.1$, $868.3$, and $868.5$ MHz, and four testing nodes attempt to disturb its transmission. These four devices will continuously transmit plain LoRa-encoded packets with a payload of 16 bytes every $2.5$ ms on four neighboring, partially overlapping frequencies and bandwidths: $868.0$, $868.2$, $868.4$, and $868.6$ MHz. 
	
	Through in-vitro efficacy experiments, we first select the most efficient plain LoRa transmitter configuration. We focus on CSS-generated signals only due to the high robustness of LoRaWan's CSS-modulated signals to any other radio signal~\cite{StaniecKowal2018, GoursaudGorce2015}. Altering the spread factor and code rate, we obtain the results of Figure~\ref{fig:two_B3tune}. The top figure illustrates how SF12 settings have a greater impact on airtime and, consequently, the retransmission count. The bottom figure confirms this, showing that the SF12-based signals induce a higher RSSI loss than the SF7. Both code rates, 4/5 and 4/8, present a similar effect. However, the more significant impact of the CR4/8 coded signal on the SF7 tests is a sign that we are mainly dealing with LoRaWAN header collision-inflicted loss, which, independently of the $4/5$ coded payload, is always coded in $4/8$~\cite{Haxhibeqirietal2017}. Thus, we use SF12 and CR4/8 as the default for the following interference tests.
	
	\begin{figure*}[tb]
		\centering
		\begin{subfigure}[t]{0.30\linewidth}
			\centering
			\begin{tikzpicture}
				\begin{axis}
					\addplot
					table[x index=0,y expr=\number\thisrowno{1} * \thisrowno{2},col sep=semicolon] 	{plotData/prelim_B3_16.csv}; 
					\addplot
					table[x index=0,y expr=\number\thisrowno{1} * \thisrowno{2},col sep=semicolon] 	{plotData/B3_16.csv}; 
					\addplot
					table[x index=0,y expr=\number\thisrowno{1} * \thisrowno{2},col sep=semicolon] 	{plotData/B3_16_TX2.csv}; 
					
					\legend{IV-Rx1, OS-Rx1, OS-Rx2}	
				\end{axis}
			\end{tikzpicture}
			\caption{Close-by, on-site vs in-vitro}
			\label{fig:two_B3cmpa}			
		\end{subfigure}
		\begin{subfigure}[t]{0.30\textwidth}
			\centering
			\begin{tikzpicture}
				\begin{axis}[%
					ylabel=RSSI,
					y unit=dBm,
					]
					\addplot
					table[x index=0,y expr=\number\thisrowno{10} * \thisrowno{2},col sep=semicolon] 	{plotData/prelim_B3_16.csv}; 
					\addplot
					table[x index=0,y expr=\number\thisrowno{10} * \thisrowno{2},col sep=semicolon] 	{plotData/B3_16.csv}; 	
					\addplot
					table[x index=0,y expr=\number\thisrowno{10} * \thisrowno{2},col sep=semicolon] 	{plotData/B3_16_TX2.csv}; 			
					
				\end{axis}
			\end{tikzpicture}
			\caption{Down-link receiver strength for Fig.~\ref{fig:two_B3cmpa}}
			\label{fig:two_B3cmpb}
		\end{subfigure}
		\begin{subfigure}[t]{0.30\textwidth}
			\centering
			\begin{tikzpicture}
				\begin{axis}[%
					legend style={at={(0.85,0.60)},
						anchor=south}
					]
					\addplot
					table[x index=0,y expr=\number\thisrowno{1} * \thisrowno{2},col sep=semicolon] 	{plotData/B3_16.csv};
					\addplot
					table[x index=0,y expr=\number\thisrowno{1} * \thisrowno{2},col sep=semicolon] 	{plotData/B3_16_TX2.csv};
					\addplot
					table[x index=0,y expr=\number\thisrowno{1} * \thisrowno{2},col sep=semicolon] 	{plotData/D2_16.csv}; 
					
					\legend{close-by-Rx1, close-by-Rx2, 10mt-Rx1}	
				\end{axis}
			\end{tikzpicture}
			\caption{Close-by vs. 10mt, on-site}
			\label{fig:two_D2cmpa}			
		\end{subfigure}
		\caption{Effect of close-by or 10 m distant interfering LoRa signals and changing downlink receive window, 16 bytes.}
		\label{fig:two_B3cmp} 
	\end{figure*}
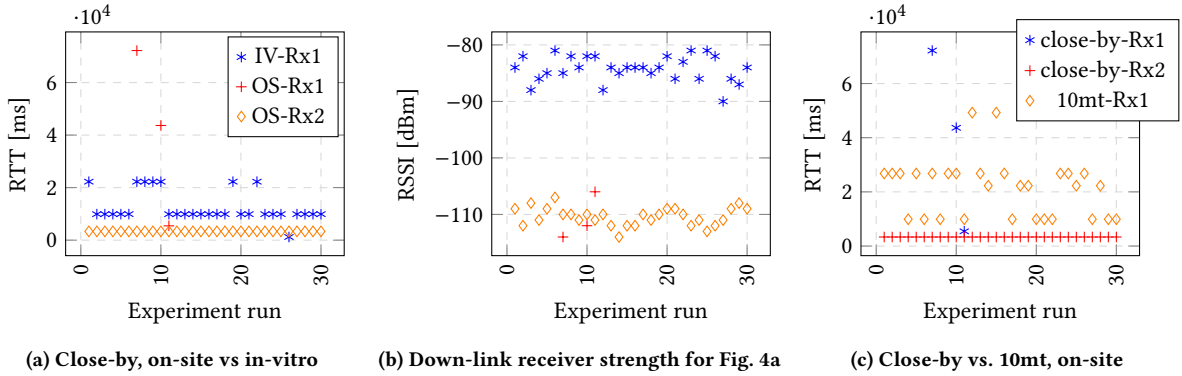	
	
	\subsubsection{Indirect interference} 
	In Figure~\ref{fig:two_B3cmp}, we plotted the results of the in vitro direct interference experiment in blue. Except for one run, all measurements show retransmission attempts and longer RTT. Most of the transmissions repeated on-site do not reach the destination. Despite multiple downlink attempts reported in the server log, only 3 (out of 10) transmissions finally reach the end node. These results converge with the tuning experiments of Figure~\ref{fig:two_B3tuneb}, displaying how a specific transmitter can reduce the RSSI of incoming signals by $15dBm$. If this interference is sufficient to reduce the signal strength below LoRaWAN's sensitivity level, it will result in transmission loss. Due to the higher distance to a gateway, the on-site location already presents downlink signal strength levels around $-110dBm$. A further test using the second receiver window (Rx2) ends with a perfect score, as shown in Figure~\ref{fig:two_B3cmpb} (orange). Such a transmitter can thus disrupt any LoRaWAN downlink transmission for a node, stalling the sending of confirmed messages.
	
	\subsubsection{Indirect interference at a distance}
	To confirm this conclusion, we set up an experiment in which the two test boxes, one acting as an interferer and one for measurement, are placed $10m$ apart. We put the latter relatively closer in line of sight to the main gateway (Gratsch, in Figure~\ref{fig:gw_locations}). Figure~\ref{fig:two_B3cmp} compares the previous close-by Rx2 window test with the system's performance with distanced test units. The results show that a $ 10ms$ distance is sufficient to bring an almost total loss back to RTTs between $10$ and $50s$, with no loss. Furthermore, the server logs reveal that other gateways also receive the uplink telegrams. The SNR level of the downlink now reaches an acceptable average of $-8dB$, allowing LoRaWAN's resilient CSS modulation to decode the downlink packets. Therefore, although it might be possible to interfere with a single node or gateway, we can not use this attack pattern for an attack on a larger scale.

	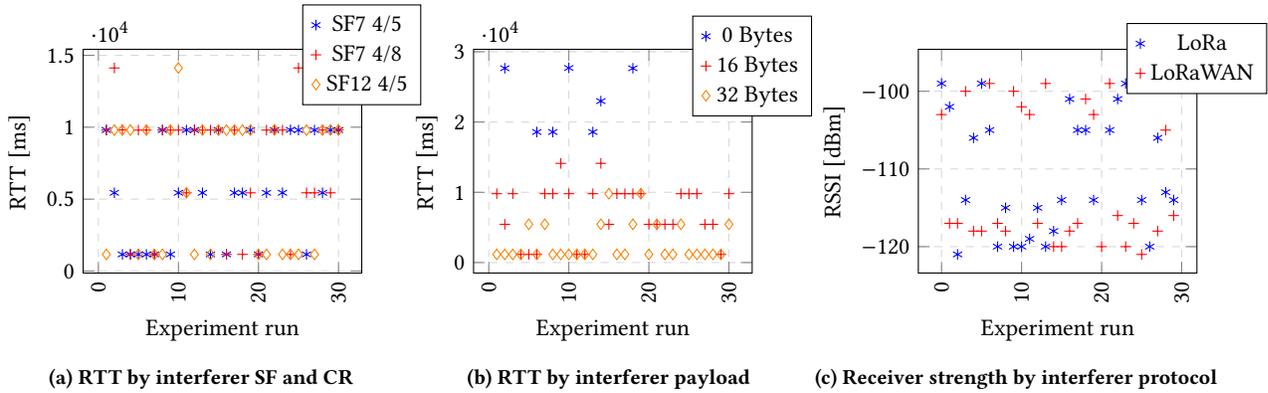
\begin{figure*}[tb]
		\centering
		\begin{subfigure}[t]{0.30\textwidth}
			\centering
			\begin{tikzpicture}
				\begin{axis}[%
					legend style={at={(1,0.75)},
						anchor=south}
					]
					\addplot
					table[x index=0,y expr=\number\thisrowno{1} * \thisrowno{2},col sep=semicolon] 	{plotData/prelim_C1_16_SF7CR5.csv}; 
					\addplot
					table[x index=0,y expr=\number\thisrowno{1} * \thisrowno{2},col sep=semicolon] 	{plotData/prelim_C1_16_SF7CR8.csv}; 
					\addplot
					table[x index=0,y expr=\number\thisrowno{1} * \thisrowno{2},col sep=semicolon] 	{plotData/prelim_C1_16_SF12CR5.csv}; 					
					
					\legend{SF7 4/5, SF7 4/8, SF12 4/5}	
				\end{axis}
			\end{tikzpicture}
			\caption{RTT by interferer SF and CR}
			\label{fig:three_C1tunea}			
		\end{subfigure}
		\begin{subfigure}[t]{0.30\textwidth}
			\centering
			\begin{tikzpicture}
				\begin{axis}[%
					legend style={at={(1,0.7)},
						anchor=south}
					]
					\addplot
					table[x index=0,y expr=\number\thisrowno{1} * \thisrowno{2},col sep=semicolon] 	{plotData/prelim_C1_0_SF7CR5.csv}; 
					%				\addplot
					%				table[x index=0,y expr=\number\thisrowno{1} * \thisrowno{2},col sep=semicolon] 	{plotData/prelim_C1_1_SF7CR5.csv}; 
					\addplot
					table[x index=0,y expr=\number\thisrowno{1} * \thisrowno{2},col sep=semicolon] 	{plotData/prelim_C1_16_SF7CR5_2.csv};
					\addplot
					table[x index=0,y expr=\number\thisrowno{1} * \thisrowno{2},col sep=semicolon] 	{plotData/prelim_C1_32_SF7CR5.csv};
					
					\legend{0 Bytes, 16 Bytes, 32 Bytes}	
				\end{axis}
			\end{tikzpicture}
			\caption{RTT by interferer payload}
			\label{fig:three_C1tuneb}
		\end{subfigure}
		\begin{subfigure}[t]{0.30\textwidth}
			\centering
			\begin{tikzpicture}
				\begin{axis}[%
					legend style={at={(1,0.80)},
						anchor=south},
					ylabel=RSSI,
					y unit=dBm
					]
					\addplot
					table[x index=6,y expr=\number\thisrowno{4},col sep=semicolon] 	{plotData/C1_255_S7C5wPr.csv}; 
					\addplot
					table[x index=6,y expr=\number\thisrowno{4},col sep=semicolon] 	{plotData/C2_16.csv}; 	
					
					\legend{LoRa, LoRaWAN}									
				\end{axis}
			\end{tikzpicture}
			\caption{Receiver strength by interferer protocol}
			\label{fig:three_C2cmpb}
		\end{subfigure}	
		\caption{End node obstruction efficacy with varying SF and CR (a), varying payload (b), or interferer protocol (c).}
		\label{fig:three_C1tune}
	\end{figure*}
	
	\subsubsection{Tuning direct interference}
	Next, we aim to understand how an adversary can affect uplink signals. For this purpose, we perform the following tests, primarily for unconfirmed transmissions, i.e., those that are uplink only. A set of LoRa efficacy tests helps to determine the optimal setting to block LoRaWAN channels if the channel frequency is known. We tested different data lengths, payload coding for SF and CR, and physical parameters, including preamble length, CRC, and header. 
	
	Figure~\ref{fig:three_C1tune} shows the results of the on-site tuning tests. The left-hand figure illustrates how a code rate of $4/8$ affects transmission more significantly than a code rate of $4/5$. At the same time, a higher spread factor shows no influence. The middle figure illustrates that shorter payloads have a more significant impact on the RTT. Both facts reflect the findings of Haxhibeqiri et al.(2017)~\cite{Haxhibeqirietal2017}, which reveal that a collision with the LoRaWAN header causes a $100\%$ transmission loss, where clashes with its payload may still show a $90\%$ success rate. Thus, for the highest impact, we want to keep the transmission as short as possible (0 bytes, SF7) while maintaining the coding rate of the header ($CR = 4/8$) and possibly omitting a preamble to shorten Tx gaps. Furthermore, we will manually set the transmission power to hardware maximum to better represent an attacker, i.e., $20dBm$.
	
	\begin{table}[tb]
		\centering
		\caption{Unconfirmed uplink statistics at the LorIoT network server with targeted 0-byte LoRa interference @$20dBm$}
		\begin{tabular}{l c c c c}
			\toprule
			Data length [bytes] & 1 (ref.) & 1  & 16  & 51 \\
			\midrule
			Receive count [x/30]& 30& 2 & 2 & 1 \\
			Best signal [dBm/dB]& -96/7.2& -102/-2.5 & -97/-2.5 & -101/-1 \\
			\bottomrule
		\end{tabular}
		\label{tab:three_interf}
	\end{table}
	
	\subsubsection{Intentional obstruction}
	\label{sec:intentional_obstruction}
	We used the eight testing nodes to attempt to interrupt the uplink transmission, one per physical channel, and measure the effect on the end device while it is transmitting with the three data lengths mentioned above. The configured setup concludes so efficiently that almost no packet reaches its intended destination, as shown in Table~\ref{tab:three_interf}. With the set configuration, the interfering test boards send a message every millisecond, disturbing almost every uplink. Interestingly, the RSSI signal is not consistently affected, while the SNR ratio is. Again, a targeted interference successfully blocks a resource, this time the uplink.
	
	\subsubsection{Channel exhaustion} verifies the efficacy of an attack using only LoRaWAN for the exhaustion of a physical channel. We repeat the previous test, setting the testing devices to send a 16-byte Smart-Lighting telegram instead of a plain LoRa message. This results in a rate of approximately 100 messages per minute (MPM) or 15-minute updates for 1,500 or 12,000 Smart-Lighting counters, respectively, at eight configured frequencies. Registered test nodes generate regular ADR adjustment downlinks, increasing overall traffic. To compare performance with LoRa messages of the same data length, we plot the signal strength of both results in Figure~\ref{fig:three_C2cmpb}. The end node achieves a delivery rate of nearly 100\% with similar signal strength in both cases. These outcomes suggest that both interferers behave similarly. However, LoRaWAN does not allow the omission of the header and CRC and is thus unable to achieve the effect seen in Section~\ref{sec:intentional_obstruction}.
	
	\subsubsection{Gateway exhaustion} 
	This test fakes rightful preambles using plain LoRa transmitters to exhaust the decoding resources of the gateway. We adjust the transmission parameters to match the preamble of LoRaWAN frames and repeatedly send only LoRaWAN preambles with each test node on one of the eight channels, attempting to interfere with the gateway's decoding process and locking up all decoding resources. Despite the flood of preambles, the gateway can manage incoming messages. For example, with 16 bytes of uplink transmissions, only 6 out of 30 communications get lost. Transmitting one or 51-byte payloads on the end node achieves a similar performance. The gateway has been appropriately configured and successfully discards the interfering packets.
	
	\subsubsection{Network exhaustion} attempts to flood the join server with requests in an attempt to deplete resources. Each testing node will continuously send OTAA or ABP "join" requests using either valid or fake keys on the three join frequencies, while the end node will use periodic OTAA-joins and confirmed messages for both tests. The results display unexpected outcomes. A repeated OTAA-join with invalid credentials does not affect the end node's transmission. When using valid OTAA credentials, however, an OTAA-join flood causes the network server to shut down the gateway, likely in an attempt to prevent server overload. An ABP-join with valid or invalid credentials effectively disrupts all communication. The server logs further indicate that, although all uplink telegrams are received, no downlink message reaches the destination before the retry count is exhausted. Thus, an ABP-join-flood has the same disruptive effect as an interfering transmission seen in \textit{indirect interference}, i.e., the content of the interfering message makes no difference. As ABP recentials are distributed on a commissioning basis, an attacker could record multiple nodes' transmissions and replay them at a later moment, potentially disabling the network for an extended period~\cite{Yangetal2018}. However, the ABP-join test showed that the frame-counter weakness has been removed in LoRaWAN specification 1.0.4~\cite{Lora2020-2}, requiring ABP counter persistence throughout a device's lifetime.
	
	\subsection{On network recovery}
	\label{sec:four}
	\begin{figure}[tb]
		\centering
		\includegraphics[width=0.6\linewidth]{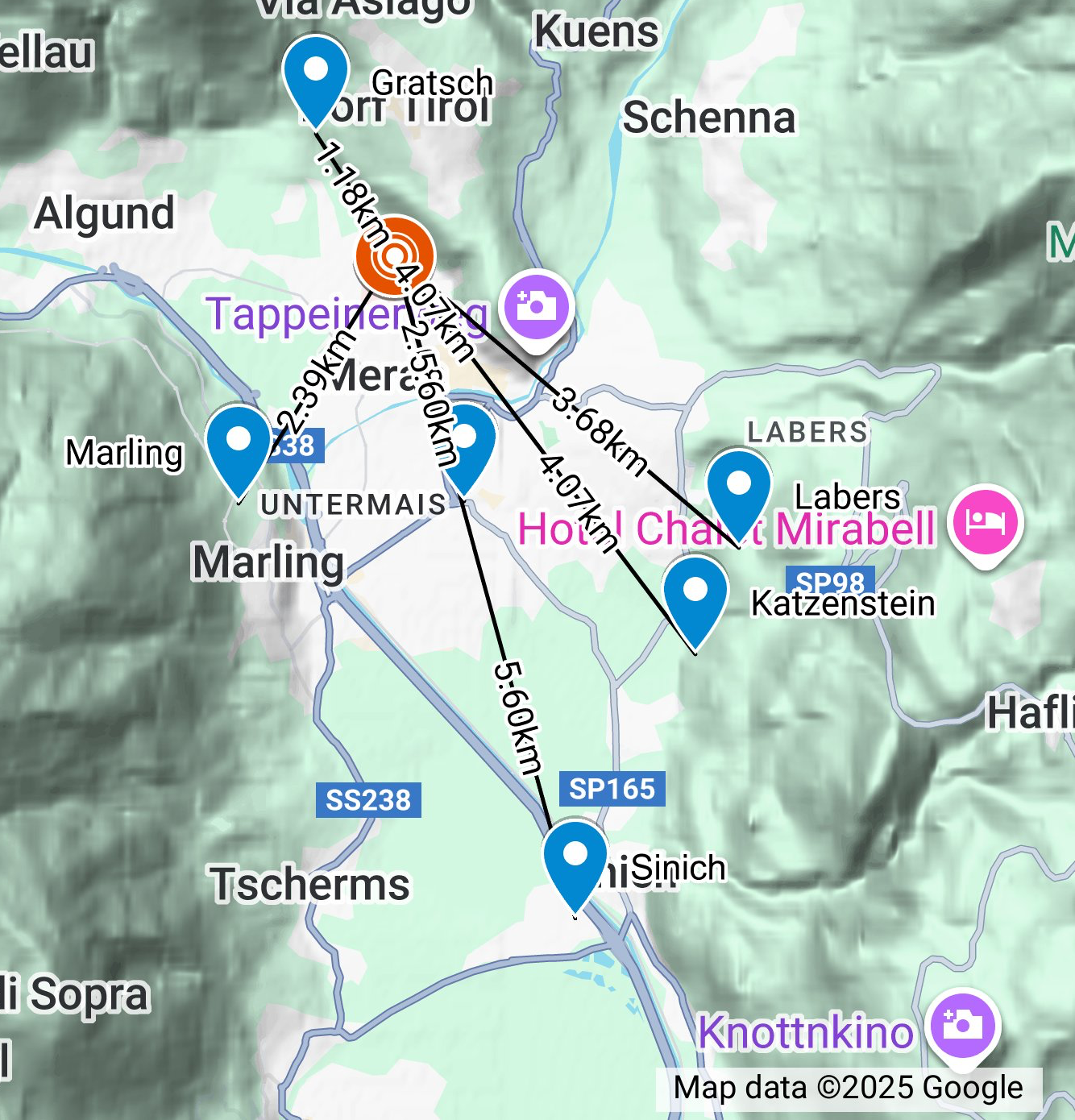}
		\caption{Gateway and TestBox locations, approx. distances.}
		\label{fig:gw_locations}
	\end{figure}
	
	For this batch, we sample the network's behavior during nightly operation. We aim to identify its response to typical traffic and to parameterize it to increase resilience, even when subject to interference or attacks. First, we observe the regular operation of the deployed network in its initial architecture configuration. Determining factors include how communication traffic begins and evolves once night falls, on both workdays and weekends. Next, we repeat the logging with an active interference, an attacker, at a location between Gratsch, Marling, and Labers Gateways; see Figure~\ref{fig:gw_locations} -- Untermais is mounted too low to receive all traffic. In this setting, the test boxes constantly transmit interfering LoRa signals on all eight frequencies at a rate of $1/ms$.
	Finally, we analyze the results and attempt to take corrective action by adjusting system or architecture parameters repeatedly until we reach the set criterion.
	
	A 24-hour traffic log collects statistics based on each lamp's three message types: boot, status, and counters. The values we want to observe are (a) the number of lamps we received each message type from and (b) their success rate or performance. We measure boot message performance via the duration between the arrival of the first and last boot messages of all lamps, thereby indirectly measuring the total boot time. Our other performance metrics compare the number of received messages to the expected messages for the particular lamp model. Finally, to assess the potential increase in the network's resilience level, we examine gateway redundancy and message coverage by measuring the number of data points covered by at least two gateways, either per device or message.
	
	\subsubsection{Default Configuration}
	Analyzing the performance of the first three nights with the default configuration, as shown in Table~\ref{tab:four_long}, we can immediately make three observations: the test boxes' interference affected the network, the received statistics were low, and the device behavior did not align with the protocol specifications. 
	
	The first is visible in the total boot time, which has jumped from approximately three minutes to almost two hours. The boot message is the first message after a device's OTAA join and can thus represent the time spent notifying the server of a device's power-on. The suffered delay thus demonstrates the effectiveness of the attack on a default-configured architecture. A quick review of the log explains the second observation. Due to system limitations, in the deployed system, a lighting controller is powered only when the lamp is in operation. Since the controllers do not store status when de-energized, those 25 configured to send data once a day will never reach the trigger for counter-transmission; an instance \textit{incorrect device configuration - $v_d$}. Lastly, in none of the initial experiments did the server receive all boot information messages from the lamps, resulting in unusually high data loss. Both these message types are confirmed, which means that the lamp should retry multiple times to resend the information. Although we expect nearly $100\%$ coverage for boot and status information across all three days, even without interference, the unlikely case of at least $5-12\%$ lost messages occurred.
	
	\begin{table}[tb]
		\centering
		\caption{24h on-site long-term network tests with 81 Lamps, number of devices and message coverage for each metric.}
		%		\small
		\begin{tabular}{l c c c |  c c | c |  c c}
			\toprule
			& \rotatebox{90}{Default} & \rotatebox{90}{Def. weekend} & \rotatebox{90}{- w/ attacker} & \rotatebox{90}{SF range ADR 6} & \rotatebox{90}{- w/ attacker} & \rotatebox{90}{Fixed SF7} & \rotatebox{90}{Fixed \& ADR 1} & \rotatebox{90}{- Weekend} \\
			\midrule
			Boot\footnotemark & 72 & 75 & 77* & 75 & 80* & 81 & 76 & 73\\
			- time [s]& 201 & 229 & 6321 & 212 & 732 & 206 & 212 & 237 \\
			Status\footnotemark[\value{footnote}] & 81 & 80 & 80* & 81 & 80* & 81 & 81 & 81\\
			- [\%] & 90.0 & 89.7 & 87.8 & 93.0 & 86.5 & 94.7 & 96.0 & 97.2 \\
			Counter & 54 & 56 & 56 & 56 & 56 & 56 & 56 & 56 \\
			- [\%] & 87.4 & 89.4 & 92.0 & 91.7 & 87.3 & 95.0  & 96.9 & 96.0\\
			GW 2+ & 7 & 17 & 13 & 12 & 10 & 14 & 17 & 14\\
			- [\%] & 87.1 & 90.9 & 83.5 & 82.5 & 73.7 & 81.3 & 81.9 & 81.5 \\
			\bottomrule
		\end{tabular}
		\label{tab:four_long}
	\end{table}	
	
	After further investigation, we found that during all three days, no lamp ever explicitly requested a missing confirmation message, thus not adhering to the LoRaWAN protocol. The extract below illustrates how a lamp controller ignores a missing downlink confirmation and the newly set data rate sent at the second $33$ (five additional data bytes); thus, an occurrence of \textit{software bugs - $v_e$} is evident. For our statistics, this means that all three message types are, unfortunately, behaving like unconfirmed messages.
	
	{
		\fontsize{6.5}{0}
		\begin{verbatim}
			03:15:33.177 DATA_UP_CNF | [..], "sf":12, [..]
			03:15:33.378 ADR DR0 => DR5 | "deveui":"70B3D5B020038857"
			03:15:33.392 DATA_DOWN | [..], "datr":"SF12BW125", "len":17, [..]
			04:15:07.128 DATA_UP_CNF | [..], "sf":12, [..]
			04:15:07.338 DATA_DOWN | [..], "datr":"SF12BW125", "len":12, [..]
		\end{verbatim}
	}
	\footnotetext{Experiments with * missed one lamp due to electrical failure. We corrected the performance values to assume the missing device was 100\% operational.}
	
	This malfunction notwithstanding, it brings up another phenomenon. While the amount of status and counters received remains similar in all three settings, the interference signal causes the status coverage to drop while the counter coverage increases. This shift in performance is likely due to the same reduction in broken status messaging. Although we described the malfunctioning confirmed telegrams as behaving like uplinks, they still trigger downlink transmissions, occupying channel resources. A successful interference in their uplink halves the time used for this kind of message, consequently reducing the risk of collision. Such a case is unlikely to happen in a functioning network. 
	
	The last statistic we collected is coverage numbers. Device coverage indicates the number of devices that have continuously operated with two or more gateways, while gateway coverage assesses the percentage of communication that reached at least two gateways. Interestingly, those values increase on the weekend. The introduced interference then reduces it again, but not to the level we observed on a weekday. This variability could be due to workday interference, as we are working on a free ISM band, general unpredictability, thus failing determinism, or simply a coincidence. 
	
	\subsubsection{Adjusted data rates}
	Observing the status and configuration on the network server, we note that most lamps barely reached a spreading factor of SF8 (DR4) despite maintaining a consistently high signal quality. The server's ADR controller should adjust these to match a device's performance. By default, the used server's algorithm adjusts the device's data rate between DR0 and DR5 (SF12 to SF7) via a piggyback message every 12 uplinks, depending on the signal quality. Because a higher SF creates prolonged airtimes, we perform the second experiment with a limited SF range. The server now adjusts the data rate every six uplinks between a minimum of DR3 (SF9) and a maximum of DR5 (SF7).
	
	Table~\ref{tab:four_long} columns four and five show the results of the SF-limited experiments. The adjusted data rates show improvements in the message coverage statistics due to shorter airtime and, consequently, lower collision probability reaching values beyond $90\%$. However, with interference, their performance is worsened. Transmissions with lower spread factors are more sensitive to interference and have, thus, an increased RSSI and SNR threshold for signal decoding, which also affects the maximum range. Indeed, the message and device gateway coverage shrank in both cases. However, comparing the two performances based on message delivery, the configuration with adjusted parameters remains favored due to its greater stability and overall performance in traffic-intensive situations.    
	
	At the end of the first night, the network server's device status logs show an overall improvement in spread factor values. All lamps with 15, 20, or 30-minute messaging periodicity reached the highest data rate, while those set to 60 minutes stopped at SF9. Indeed, with an ADR counter setting of 6 uplinks and a 60-minute messaging periodicity, the server makes only one attempt to adjust the data rate in a $10h$ nightly operation. However, this still does not explain the slow reactivity of the lamp posts in adjusting their data rate. 
	
	Examining the device logs, we observe two anomalies during the lamp's power-up sequence. First, after successfully joining the network using SF7 (DR5), the device starts communicating in SF12 (DR0) despite our imposed data rate and a successful join at the highest data rate. Second, the log presents two join requests within a few seconds; see below. Join requests have a longer Rx1 and Rx2 window, the latter opening $6s$ after transmission. Considering a random retransmission delay range of 1 to 3 seconds for all messages requiring a server response, the device should not initiate a second attempt earlier than $7s$ after the first failed request. However, aside from the unsynchronized gateway clocks, the two incoming joins are less than $1s$ distant, violating the prescribed back-off timeout of the LoRaWAN protocol. However, it may also be an instance of \textit{inconsistent and incomplete specifications - $v_g$}. While it is true that back-off timeout is mandatory for all confirmed messaging, one can still argue that a \texttt{Join\_Request} is not such a type of message. Likewise, the DR to set after a join is not unequivocally specified.
	
	{
		\fontsize{6.5}{0}
		\begin{verbatim}
			20:18:56.005 JOIN_REQUEST | "sf":7, "gweui":"024B0BFFFF030FF5",
			"time":"18:19:02.855550927Z", "devnonce":51313, [..]
			20:18:56.035 JOIN_ACCEPT | [..], "datr":"SF7BW125", [..]
			20:18:56.913 JOIN_REQUEST| "sf":7, "gweui":"024B0BFFFF030703",
			"time":"18:18:56.632556000Z", "devnonce":57973, [..]
			20:18:56.921 JOIN_ACCEPT | [..],"datr":"SF7BW125",[..]
			20:21:54.657 DATA_UP_CNF | [..], "seqno":0, "sf":12, [..]
		\end{verbatim}	
	}
	
	\subsubsection{Fixed data rates}
	Considering these outcomes and the set of discovered problems that hinder the network from operating at full throughput, we conducted the third experiment set with a fixed data rate of DR5 (SF7) and the fourth with ADR trigger counters set to one. The results demonstrate that setting the data rate to DR5 (SF7) as soon as possible significantly improves the total network throughput; see Table~\ref {tab:four_long}, last three columns. Specifically, shortening the ADR trigger time caused an increase of $1.5-2\%$ in message coverage. The device coverage of the gateways stays at high levels; however, as expected, the gateway message coverage slightly reduces compared to a non-fixed SF setting due to the lower reach of higher data rates. -- 97\% coverage reached.
	
	%With coverage reaching the target of 97\% despite these protocol faults, we stop the experiment iteration at this point.
	
	\section{Takeaways for practitioners}
	\label{sec:discussion}
	
	\textit{Resource vulnerabilities - $v_a$} are present but not significant. The network's robustness against interference appears remarkable. High sensitivity and the ability to decode messages received with a negative SNR make the network resilient to any CSS-based interference with a transmission power weaker than the signal being processed. Additionally, the decoder can even handle interfering non-LoRa signals between $ 5$ dBm and $ 19.5$ dBm stronger than the data to be decoded, depending on the spread factor~\cite{GoursaudGorce2015}. Furthermore, the data shows that most retransmissions are due to missing downlink reception. The confirmed message was usually not interrupted by an actual uplink collision but rather by interference through an attacker's signal with a higher RSSI than the downlink. As confirmed transmissions are rather rare in an SLA, the effect of an attack is thus limited.
	
	Resource vulnerabilities involving gateways and join servers were also less effective than expected. The gateway was able to buffer incoming payloads sufficiently to process and discard most telegrams. Even though it is unclear whether the lost uplinks are due to gateway overload or colliding traffic, the network demonstrated sufficient robustness to such attacks, with approximately $30\%$ data loss. Likewise, the OTAA join floods did not affect the network using fake credentials. The only successful attacks were ABP join-floods that behave like a disturbing signal or valid OTAA join-floods that cause the system to shut down the gateway. However, if proper off-band security protocols are respected, the latter attacks should be no concern.
	
	An adversary close to a node attempting to exploit \textit{large-scale ($v_b$)} and \textit{wide distribution ($v_c$)} incurs additional obstacles. As lamp posts are typically many meters apart, the effect of an interfering signal with power up to $20dBm$ remains limited. Indeed, we have seen how a distance of $10m$ is already enough to weaken the effect of interference, which, if set close by, impedes almost all communication. Therefore, positioning an interference signal close to a gateway would achieve a greater impact. However, a second gateway will intercept and decode the payload once a network grants redundancy, nulling the interference effect. If a device is compromised and credentials are stolen, those could be effectively used to disable gateways throughout the system, posing a valid threat. Physical node access must thus be hindered as much as possible. 
	
	We investigated weaknesses and changes that would increase the system's resilience and recovery performance during the second batch. Among the weaknesses, we identified multiple incorrect implementations of the LoRaWAN protocol ($v_d$, $v_g$), accompanied by erroneous parameterization and system usage ($v_f$). The confirmed-uplink malfunction and the abnormal behavior of the data rate control demonstrate the importance of thoroughly testing the end nodes \textit{before} installation. Due to these software errors, the network will never reach its full performance. While it is possible to configure a network to function even under such adverse circumstances, the required effort to restore proper operation tends to be high.
	
	\section{Conclusions}
	\label{sec:conclusions}
	
	This paper presents experiments conducted to evaluate the cybersecurity of a LoRaWAN network. We designed stress tests to identify vulnerabilities identified in prior analyses and found that most were either absent or not severe enough to significantly disrupt the network. Long-term monitoring and data analysis were employed to enhance system robustness against attacks, despite potential interference. Log analysis revealed several specification violations in the LoRaWAN lighting controllers, leading to network congestion due to failures in adhering to back-off times and retransmission requirements. We also proposed additional measures to bolster network resilience. Future work will focus on analyzing the network's growth with more gateways and over 1000 nodes, and exploring how AI-supported parameter optimization could benefit such a dense network.
	
	\begin{acks}
		The authors would like to thank the Municipality of the City of Merano and \emph{systems s.r.l.} for providing the infrastructure and support. The work has been funded by project no. EFRE1039 in the EFRE-FESR 2021-2027 program and by the project PRIN 2022 PNRR BeT - codice 2022TEPX4R CUP I53D23003730008 Finanziamento dell’Unione Europea – NextGenerationEU – PNRR  M4.C2.1.1.
	\end{acks}
	
	\bibliographystyle{ACM-Reference-Format}
	\bibliography{bibliography}
	
\end{document}